\documentclass{sty/IEEEtran}

\usepackage{amsmath}
\usepackage{amsfonts}
\usepackage{amssymb}
\usepackage{graphicx}
\usepackage{booktabs}
\usepackage{siunitx}
\ifCLASSOPTIONcompsoc
\usepackage[caption=false, font=normalsize, labelfont=sf, textfont=sf]{subfig}
\else
\usepackage[caption=false, font=footnotesize]{subfig}
\fi

\usepackage{cite}

\usepackage{threeparttable}
\usepackage{multirow}
\usepackage{times}
\usepackage{float}
\usepackage{enumerate}
\usepackage[none]{hyphenat}
\usepackage[normalem]{ulem}
\usepackage{color}
\usepackage{xcolor}
\usepackage{url}

\usepackage{tikz}
\usepackage{pgfplots}
\pgfplotsset{compat=newest}
\usetikzlibrary{positioning}

\usepackage{algorithm}
\usepackage{algorithmic}

\newcommand{\qed}{\nobreak \ifvmode \relax \else
	\ifdim\lastskip<1.5em \hskip-\lastskip
	\hskip22em plus0em minus0.5em \fi \nobreak
	\vrule height0.4em width0.3em depth0.25em\fi}

\newlength\fheight
\newlength\fwidth
\begin{document}
	\title{Fast Angular Sweep Monostatic Radar Cross Section Computation via Model Order Reduction}
	\author{Noelia Naranjo and Valent\'{i}n de la Rubia
	\thanks{Noelia Naranjo and Valent\'{i}n de la Rubia are with the Departamento de Matem\'{a}tica Aplicada a las {TIC}, ETSI de Telecomunicaci\'{o}n, Universidad Polit\'{e}cnica de 
	Madrid, 28040 Madrid, Spain (e-mails: noelia.naranjo@upm.es; valentin.delarubia@upm.es).}
	}
	\maketitle
	\begin{abstract}
		The radar cross section (RCS) is a fundamental parameter in radar engineering, as it determines the detectability of a given target by a radar system. Increasing interest in very low observability (VLO) targets has arisen recently. As a result, especial effort is put in developing numerical tools to accurately predict the electromagnetic scattering from perfect electric conductor objects. Integral equation methods are commonly used in this task. However, full-wave computation of the monostatic RCS at a single angle can be rather time-consuming, nevermind computing the RCS in a specific angular interval with a fine sampling. In fact, this is what is needed in industrial applications.
		
		A reduced-order model for fast full-wave monostatic RCS evaluation is proposed to easily achieve fine details in the desired angular domain. An open source integral equation solver, taken into account as a black box, is used to get the electromagnetic scattering in non-penetrable objects at some specific incident angles. Several radar targets will show the possibilities and capabilities of these model order reduction approaches.
	\end{abstract}
	%
	%
	\begin{keywords}
		Computational electromagnetics (CEM), computational prototyping, electromagnetic scattering, integral equation methods, model order reduction, method of moments, radar cross section.
	\end{keywords}
	\IEEEpeerreviewmaketitle
\section{Introduction}
\label{Sec-Introduction}
\PARstart{T}{he} radar cross section (RCS) is a measure of power scattered in a given direction when a target is illuminated by an incident wave. It specifically indicates how detectable an object is by radar. RCS is defined to characterize the properties of the target, rather than the effects of transmitter power, receiver sensitivity or the distance between transmitter and receiver. This parameter is essential in fields such as object detection, aircraft and vehicle design with stealth characteristics, or the evaluation of communication and defense systems \cite{MattRCS2010,JinFaRCS2023,LandesaRCS2025}.

The numerical computation of RCS is a highly demanding problem. One of the most widely used techniques is the Method of Moments (MoM), which provides accurate solutions of integral equations describing the interaction between incident waves and induced surface currents on the object. However, its main drawback lies in the high computational complexity it entails, particularly when analyzing complex geometries or performing simulations for a wide range of frequencies and incidence angles. This limitation reduces its practical applicability in scenarios where efficiency and computational cost are critical. As a matter of fact, we are specifically interested in getting the scattered field by a PEC target for multiple incident angles, which implies solving the integral equation for different right hand sides, i.e., excitation vectors. To address this issue, either a direct inversion of the MoM matrix or an iterative solution for multiple right hand sides can be carried out. Either approach have their own pros and cons. Among fast direct solver approaches, we refer to \cite{Michielssen2017,Rius2026}. Notably, we draw the attention to the recent work \cite{Rius2026}, where a fast $N \log^4 N$ complexity matrix factorization is detailed. Among iterative approaches, which have traditionally kept much of the attention, the works in \cite{MattRCS2010,Schuster2012,Sheng2014,JinFaRCS2023,LandesaRCS2025} are highlighted. These so-called iterative approaches need to deal with the multiple right hand side (multiple excitation vector) solutions in a smart way, either applying compressed sensing, linear algebra, or even Fourier analysis. By the same token, another interesting approach that has been kept under the radar is model order reduction. Model order reduction approaches \cite{benner2015survey,Silveira2014Reduced-OrderModels} make it possible to approximate the behavior of high-dimensional systems by a reduced-dimension model, namely, a reduced-order model (ROM), that keeps their main characteristics, significantly decreasing computation time while maintaining accuracy.

In this work, we use a model order reduction technique to tackle this computational challenge, called the Reduced Basis Method (RBM) \cite{de2009reliable,fares2011reducedRBMEFIE,monje2020electricRBMEFIE}. RBM approximates the solutions through the construction of a reduced dimension subspace obtained from a set of MoM solutions. These solutions in our integral equation problem are the induced currents on the perfect electric conductor (PEC) object, and we wish to get fast evaluation of all these currents as we change the incident angle of the impinging plane wave on the target. A common strategy for building this reduced basis is the use of greedy algorithms, which iteratively select the most representative components to enrich the so-called reduced basis space \cite{Edlinger2014APosteriori,morRewLM15,chellappa2023infsup,delaRubia2022fastaposteriori}. Moreover, when the model includes non-linear or non-affine parameter-dependent terms, additional strategies are required to enable efficient ROM evaluation. In this context, the Discrete Empirical Interpolation Method (DEIM) \cite{chaturantabut2010nonlinearDEIM,Silveira2014Reduced-OrderModels} provides an effective way to interpolate parameter-dependent terms efficiently, thereby reducing the number of required MoM evaluations without compromising accuracy. 

This work is organized as follows. In Section \ref{Sec-ProblemStatement}, we review the MoM formulation to stress out the difficulties to get a fast monostatic RCS prediction for a PEC object. Section \ref{Sec-RBM} discusses in detail the reduced order model approach to speed the simulation time up without sacrifying accuracy in monostatic RCS evaluation. In Section \ref{Sec-NumericalResults}, we address several challenging PEC targets, including a turbine inlet model and an F-16 aircraft model, to show the capabilities and accuracy of the proposed methodology. Finally, in Section \ref{Sec-Conclusions}, we comment on the conclusions.

\section{Problem Statement}
\label{Sec-ProblemStatement}
Electromagnetic scattering problems are modeled using Maxwell's equations in the harmonic regime. For a PEC target, the application of the Stratton–Chu representation leads to the Electric Field Integral Equation (EFIE), which reformulates the problem as a boundary integral equation defined on the surface of the scatterer. To solve it numerically, the EFIE is expressed in weak form over a suitable admissible function space $\mathcal{V}$, defined as follows: 

Find $\mathbf{J} \in \mathcal{V}$ such that
\begin{equation}
\label{variationalProblem}
    a(\mathbf{J},\mathbf{v}) = f(\mathbf{v}), \, \, \forall  \mathbf{v} \in \mathcal{V}
\end{equation}
with the bilinear form
\begin{equation}
\label{bilinearForm}
    \begin{aligned}
        a(\mathbf{J}, \mathbf{v}) = i \omega \mu\ 
        \int_{\partial \Omega} \int_{\partial \Omega}
        G(\mathbf{r} - \mathbf{r}'; k) \, 
        \mathbf{J}(\mathbf{r}') \cdot \mathbf{v}(\mathbf{r}) 
        \, ds' \, ds \\
        + \frac{1}{i\omega\varepsilon} 
        \int_{\partial \Omega} \int_{\partial \Omega}
        G(\mathbf{r} - \mathbf{r}'; k) \, 
        \nabla \cdot \mathbf{J}(\mathbf{r}') \, 
        \nabla \cdot \mathbf{v}(\mathbf{r}) 
        \, ds' \, ds 
    \end{aligned}
\end{equation}
and the linear form
\begin{equation}
    f(\mathbf{v}) = \int_{\partial \Omega} \mathbf{E^{inc}} \cdot \, \mathbf{v} \, \, dS \text{.}
\end{equation}

Here, $G(\mathbf{r} - \mathbf{r}'; k)$ is the free-space Green function and the admissible space $\mathcal{V}$ is a trace space of $H(\mathrm{curl}, \Omega)$, namely
\begin{equation}
    H^{-1/2}(\mathrm{Div}, \partial \Omega) 
    = \left\{ \mathbf{n} \times \mathbf{u} \;\; \text{on } \partial \Omega \;\big|\; \mathbf{u} \in H(\mathrm{curl}, \Omega) \right\},
\end{equation}

The variational problem (\ref{variationalProblem}) is typically solved using the method of moments (MoM), which approximates the surface current in a finite-dimensional subspace $\mathcal{V}_h \subset \mathcal{V}$ generated by Rao–Wilton–Glisson (RWG) basis functions defined on pairs of adjacent triangles in the discretized mesh of $\partial \Omega$, surface boundary of the PEC object. The unknown surface current $\mathbf{J}$ is then expressed as a linear combination of these basis functions $\{ \mathbf{v}_i \}_{i=1}^n$:
\begin{equation}
    \mathbf{J}(\varphi) \approx \sum_{i=1}^n I_i(\varphi) \mathbf{v}_i,
\end{equation}
where $I_i(\varphi) $ are the unknown coefficients to be determined. 
Substituting this expression into (\ref{variationalProblem}) and testing against the same set of basis functions leads to a discrete system of equations:
\begin{equation}
\label{MOMProblem}
    Z \, I(\varphi) = V(\varphi)
\end{equation}
where $I(\varphi) \in \mathbb{C}^{n \times 1}$ is the solution vector, $Z \in \mathbb{C}^{n \times n}$ is the EFIE MoM impedance matrix and $V(\varphi) \in \mathbb{C}^{n \times 1}$ is the excitation vector.  

Once $I(\varphi)$ is computed, an approximation of the induced current is found and, consequently, the scattered field and radar cross section can be obtained. It is important to note that the method of moments leads to a dense system matrix of size $n \times n$, where $n \gg 1$. Assembling and solving this system is computationally expensive, which motivates the use of reduced-order modeling techniques such as RBM and DEIM \cite{de2009reliable,chaturantabut2010nonlinearDEIM,fares2011reducedRBMEFIE}.

\section{Reduced Basis Method}
\label{Sec-RBM}
In practical applications, the RCS must be evaluated over a wide range of incidence angles, which requires solving problem (\ref{MOMProblem}) repeatedly, resulting in a significant computational problem. However, the variation of $I(\varphi)$ with respect to $\varphi$ is relatively smooth, suggesting that the family of solutions can be efficiently approximated in a space of much lower dimension than that of the entire problem. The Reduced Basis Method (RBM) is presented as an effective tool for this purpose. By constructing a subspace generated from a limited number of solutions $I(\varphi_i) \equiv b_i$, called snapshots, it is possible to accurately approximate the system behavior.

Using the RBM projection basis $B=cols\{b_i \equiv I(\varphi_i) \}_{i=1}^{r}$, where matrix $B \in \mathbb{C}^{n \times r}$, we can approximate the solution vector $I(\varphi)$ in (\ref{MOMProblem}) as
\begin{equation}
    \label{approximation}
    I(\varphi) \approx B \tilde{I}(\varphi) = \sum_i^r \tilde{I}_i(\varphi)b_i \text{.}
\end{equation}
By means of Galerkin projection of (\ref{MOMProblem}) onto the generated subspace, we obtain the following reduced system:
\begin{equation}
    \label{Sec-RBM-reducedEquation}
     B^TZ \, B \tilde{I}(\varphi) = B^TV(\varphi)
\end{equation}
\begin{equation}
    \label{Sec-RBM-reducedEquation2}
     \tilde{Z}\, \tilde{I}(\varphi) = \tilde{V}(\varphi) \text{,}
\end{equation}
where $\tilde{Z} \equiv B^TZ \, B \in \mathbb{C}^{r \times r}$, $\tilde{V}(\varphi) \equiv B^TV(\varphi) \in \mathbb{C}^{r \times 1}$, $\tilde{I}(\varphi) \in \mathbb{C}^{r \times 1}$ and $r \ll n$. This new reduced problem \eqref{Sec-RBM-reducedEquation2} can be solved in a much less expensive way in comparison with the MoM (\ref{MOMProblem}), saving many computational resources.

In order for the RBM approximation to be valid, \eqref{approximation} and \eqref{MOMProblem} must hold. Then, we can state that $\tilde{I}(\varphi) \in \mathbb{C}^{r \times 1}$ must be the vector that makes the following residual as close to zero as possible:
\begin{equation}
    \label{residual}
    R(\varphi) = V(\varphi) - ZB\tilde{I}(\varphi) = V(\varphi)-\sum_{i=1}^r \tilde{I_i}(\varphi)Z\,b_i \text{.}
\end{equation}

This is equivalent to minimize the norm of this residual:
\begin{equation}
	\label{Sec-ProblemStatement-eq-residualNorm}
	\begin{aligned}
    \|R(\varphi)\|^2 &= (V(\varphi), V(\varphi)) - 2 \sum_{i=1}^r \text{Re} \{\tilde{I}_i^*(\varphi)(V(\varphi),Z\,b_i)\} \\
	&+ \sum_{i=1}^r \sum_{j=1}^r \tilde{I}_i(\varphi) \tilde{I}_j^*(\varphi)(Z\,b_i,Z\,b_j) \text{,} 
	\end{aligned}
\end{equation}
where $*$ denotes complex conjugate. In this way, once we have calculated the inner products $(\cdot,\cdot)$ among vectors $Z\,b_i$, which are done once and stored in memory, \emph{evaluating the norm of this residual error for any incident angle} $\varphi$ is a computationally simple operation.

Thus, the procedure for selecting suitable reduced basis vectors $I(\varphi_i)$, and therefore, the reliability of this approximation, is based on a greedy algorithm (see Algorithm~\ref{greedy}) that minimizes the residual error $\|R(\varphi)\|$ along the angular band of study $B$. In this way, we need to solve the MoM problem only at certain $\varphi_i$  to build up the reduced-basis space, namely, we chose those $\varphi_i$ that maximize the residual norm at each loop iteration in the greedy algorithm, which can be quickly calculated by means of (\ref{Sec-ProblemStatement-eq-residualNorm}). Once we have assembled the reduced-basis projection matrix $B$, we can compute the electric current $I(\varphi)$ in the MoM discretization space at any angle in the band of interest solving the reduced order model (\ref{Sec-RBM-reducedEquation2}), which is very cheap, and using (\ref{approximation}).

\begin{algorithm}[H]
\caption{Greedy Algorithm}
\label{greedy}
\begin{algorithmic}[1]
\STATE Choose $\varphi_1$ in $\mathcal{B}$ randomly.
\STATE Solve the MoM problem (\ref{MOMProblem}) and obtain $I(\varphi_1)$.
\STATE Set $B =$ cols$\{I(\varphi_1)\}$.
\STATE Solve the reduced problem \eqref{Sec-RBM-reducedEquation2} in $\mathcal{B}$.
\STATE Compute the residual $\|R(\varphi)\|$ in $\mathcal{B}$.
\WHILE{$\|R(\varphi)\| > \epsilon_{tol}$}
    \STATE Choose $\varphi^\star = \arg\max_{\varphi \in \mathcal{B}} \|R(\varphi)\|$
    \STATE Solve the MoM problem \eqref{MOMProblem} at $\varphi^\star$.
    \IF{$\{I(\varphi^\star)\} \not \subset \text{span}\{B\}$}
        \STATE Set $B \leftarrow B \cup \text{cols} \{I(\varphi^\star)\}$.
        \STATE Solve the new reduced problem \eqref{Sec-RBM-reducedEquation2} in $\mathcal{B}$.
        \STATE Compute the residual $\|R(\varphi)\|$ in $\mathcal{B}$.
    \ELSE
        \STATE \textbf{break}
    \ENDIF
\ENDWHILE
\end{algorithmic}
\end{algorithm}

\section{Discrete Empirical Interpolation Method}
\label{Sec-DEIM}
Although the reduced basis method (RBM) significantly reduces the problem dimension for the electric current $I(\varphi)$, there is still a point that needs to be properly addressed, namely, the reduced-order model \eqref{Sec-RBM-reducedEquation}-\eqref{Sec-RBM-reducedEquation2} still contains a parameter-dependent term $V(\varphi)$ whose evaluation depends on the dimension of MoM. This point is of paramount importance since it limits the overall computational efficiency of the method, particularly in monostatic RCS analysis, where the response must be computed for many incidence angles and the excitation vector $V(\varphi)$ must therefore be evaluated repeatedly.

The Discrete Empirical Interpolation Method (DEIM) \cite{chaturantabut2010nonlinearDEIM} provides an efficient way to address this issue by combining projection and interpolation. In this context, DEIM approximates the space generated by the excitation vector $V(\varphi)$ as a function of the parameter $\varphi$ (incident angle) with a low-dimensional subspace spanned by a basis \{$u_1,..., u_m\}\subset\mathbb{C}^{n \times 1}$ with $m \ll n$. Recall $n$ is the dimension of MoM.

The DEIM approximation is given formally as follows. See \cite{chaturantabut2010nonlinearDEIM} for further details.

Let $V : \mathcal{D} \to \mathbb{C}^{n \times 1}$ be a parameter-dependent vector-valued function with 
$\mathcal{D} \subset \mathbb{R}^d$ for some positive integer $d$. 
Let $\{u_\ell\}_{\ell=1}^m \subset \mathbb{C}^{n \times 1}$ be a linearly independent set 
for $m \in \{1,\dots,n\}$. 
For $\varphi \in \mathcal{D}$, the DEIM approximation $V^\star(\varphi)$ of order $m$ for $V(\varphi)$ in the space spanned by $\{u_i\}_{i=1}^m$ is given by
\begin{equation}
    \label{hatV}
     V(\varphi) \approx V^\star(\varphi) = U (P^T U)^{-1} P^T V(\varphi)\text{,}
\end{equation}
where $U = [u_1, \dots, u_m] \in \mathbb{C}^{n \times m}$ and 
$P = [e_{\wp_1}, \dots, e_{\wp_m}] \in \mathbb{C}^{n \times m}$, 
with $\{\wp_1,\dots,\wp_m\}$ being the interpolation indices selected from the input basis $\{u_i\}_{i=1}^m$. Note that $e_{k}$ is the k-th column of the identity matrix in $\mathbb{C}^{n \times n}$.

To obtain this approximation, we must specify the projection basis ${u_1,..., u_m}$ and the interpolation indices ${\wp_1, \ldots, \wp_m}$. The projection basis \(U\) is obtained from a set of excitation snapshots evaluated at different incidence angles. Let $V_{\mathrm{snap}}$ be the snapshot matrix, namely,
\begin{equation}
    V_{\mathrm{snap}} = \left[ V(\varphi_1), V(\varphi_2), \ldots, V(\varphi_s) \right] \in \mathbb{C}^{n \times s} \text{,}
\end{equation}
where \(s\) is the number of sampled incidence angles. The choice of \(s\) is relevant for the construction of the DEIM basis, since an insufficient number of snapshots may not properly capture the angular variation of the excitation vector, whereas an excessively dense sampling unnecessarily increases the computational cost of the offline stage. Therefore, rather than selecting \(s\) arbitrarily, the angular sampling criterion proposed in \cite{lee2023efficient} is adopted. For a given target, the maximum angular step, $\Delta\phi$ or $\Delta\theta$, is estimated as
\begin{equation}
    \label{angularStepCriterion}
    \Delta\phi,\Delta\theta
    \leq
    \frac{\pi}
    {k_0D + 1.8\,d_0^{2/3}(k_0D)^{1/3}},
\end{equation}
where \(k_0\) is the free-space wavenumber, \(D\) is the diameter of the sphere that encloses the target and \(d_0\) is the number of digits of the desired accuracy. Then, we can determine the number of incident excitation vectors $s$ by
\begin{equation}
    \label{numberSnapshots}
    s =
    \left(
    \left\lceil
    \frac{|\phi_{\mathrm{end}}-
    \phi_{\mathrm{start}}|}
    {\Delta\phi}
    \right\rceil + 1 \right)
    \left(\left\lceil
    \frac{|\theta_{\mathrm{end}}-\theta_{\mathrm{start}}|}
    {\Delta\theta}
    \right\rceil + 1
    \right)\text{,}
\end{equation}
where $\lceil \cdot \rceil$ is the ceiling function.

A truncated Proper Orthogonal Decomposition (POD) is then performed by means of the singular value decomposition (SVD)
\begin{equation}
    V_{\mathrm{snap}} \approx U_m \Sigma_m W_m^H,
\end{equation}
where \(U_m \in \mathbb{C}^{n \times m}\) contains the \(m\)
dominant left singular vectors, \(\Sigma_m \in \mathbb{R}^{m \times m}\)
contains the associated singular values, and
\(W_m \in \mathbb{C}^{s \times m}\) contains the corresponding
right singular vectors. The DEIM projection basis is then defined as $U \equiv U_m = \left[ u_1, u_2, \ldots, u_m \right]$.

Equivalently, each DEIM mode can be expressed as a linear
combination of the excitation snapshots:
\begin{equation}
    \label{lineal-combination u_i}
    u_i = \frac{1}{\sigma_i} \sum_{j=1}^{s} V(\varphi_j) w_{j,i}, \qquad i=1,\ldots,m,
\end{equation}
where \(\sigma_i\) is the \(i\)-th singular value and \(w_{j,i}\)
denotes the \(j\)-th component of the \(i\)-th right singular
vector. Therefore, each column of \(U\) is a linear combination of the
excitation snapshots $V(\varphi_j)$, with coefficients determined by the singular values and right singular vectors. See \cite{QuarteroniManzoniRBM,HesthavenRozzaRBM} for the details. The interpolation indices ${\wp_1}, \ldots, {\wp_m}$ are selected inductively from the projection basis $U$ by the DEIM algorithm  (see Algorithm~\ref{deim}).

Applying the DEIM approximation $V^\star(\varphi)$ of the excitation vector $V(\varphi)$ in \eqref{hatV} to the reduced excitation term $\tilde{V}(\varphi)=B^TV(\varphi)$ we get:
\begin{equation}
    \tilde{V}(\varphi) \approx B^T V^\star(\varphi) = B^TU (P^T U)^{-1} P^TV(\varphi) \text{.}
\end{equation}

The term $B^TU (P^T U)^{-1}\in \mathbb{C}^{r \times m}$ does not depend on $\varphi$ and therefore can be precomputed before solving the system. In addition, $P^TV(\varphi) \in \mathbb{C}^{m \times 1}$ can be obtained by computing the components ${\wp_1}, \ldots, {\wp_m}$ in $V(\varphi) \in \mathbb{C}^{n \times 1}$. Note that $m \ll n$. As a matter of fact, this feature is especially important in monostatic RCS analysis, where the response must be evaluated for multiple angles of incidence. Instead of evaluating all the $n$ components in $V(\varphi)$ for each incidence angle $\varphi$, DEIM requires computing only $m$ components in the excitation vector $V(\varphi)$ corresponding to the selected indices ${\wp_1}, \ldots, {\wp_m}$.

\begin{algorithm}[H]
\caption{DEIM}
\label{deim}
\textbf{INPUT:} $\{u_\ell\}_{\ell=1}^m \subset \mathbb{C}^{n \times 1}$\\
\textbf{OUTPUT:} $\wp = [\wp_1,\dots,\wp_m]^T$
\begin{algorithmic}[1]

\STATE $[|\rho|,\; \wp_1] = \max\{|u_1|\}$ /* $\wp_1$ is the index where the max is found. */
\STATE $U = [u_1],\; P = [e_{\wp_1}],\; \wp = [\wp_1]$

\FOR{$\ell = 2$ to $m$}
  \STATE Solve $(P^T U)\,c = P^T u_\ell$ for $c$
  \STATE $r = u_\ell - U c$
  \STATE $[|\rho|,\; \wp_\ell] = \max\{|r|\}$ /* $\wp_\ell$ is the index where the max is found. */
  \STATE $U \leftarrow [U\; u_\ell], \; P \leftarrow [P\; e_{\wp_\ell}], \; \wp \leftarrow \begin{bmatrix}\wp\\ \wp_\ell\end{bmatrix}$
\ENDFOR

\end{algorithmic}
\end{algorithm}
\section{DEIM Approximation of the Far-Field Operator}
\label{Sec-FarFieldDEIM}
Once the reduced system \eqref{Sec-RBM-reducedEquation2} has been solved, the reduced coefficient vector for the electric current \(\widetilde{I}(\varphi)\) is obtained. An approximation to the MoM solution for the electric current $I(\varphi)$ can then be recovered as
\begin{equation}
    \label{approximation2}
    I(\varphi) \approx I^\star(\varphi) = B\widetilde{I}(\varphi) \text{.}
\end{equation} 
This operation reconstructs the solution in the MoM space of dimension $n$. Let $\mathcal{F}(\varphi) \in \mathbb{C}^{1 \times n}$ be the scattered far-field linear operator associated with a given observation angle \(\varphi\), the corresponding far-field quantity can be written as
\begin{equation}
    e(\varphi) = \mathcal{F}(\varphi) I(\varphi) \text{.}
\end{equation}
Using the approximation (\ref{approximation2}), this expression becomes
\begin{equation}
    \label{far-field equation}
    e(\varphi) \approx  \mathcal{F}(\varphi) I^\star(\varphi) = \mathcal{F}(\varphi) B \widetilde{I}(\varphi) \text{.}
\end{equation}
A direct evaluation of this expression would require assembling the MoM dimension electric current approximation $I^\star(\varphi)$ up for each incident angle $\varphi$. Thus, although the linear system is solved in the reduced space of dimension $r$, the scattered far-field evaluation still involves working within the MoM space of dimension $n$ ($n \gg r$). Note that $\mathcal{F}(\varphi) \in \mathbb{C}^{1 \times n}$. To avoid this step, the far-field quantity is evaluated directly from the reduced coefficients \(\widetilde{I}(\varphi)\) by introducing a DEIM approximation of the scattered far-field operator $\mathcal{F}(\varphi)$.

Since \(\mathcal{F}(\varphi)\) is a row vector, the DEIM approximation ${\mathcal{F}}^\star(\varphi)$ is applied to its transpose, \(\mathcal{F}(\varphi)^T \in \mathbb{C}^{n \times 1}\). Following the DEIM formulation introduced for the excitation vector in Section \ref{Sec-DEIM}, we write
\begin{equation}
    {\mathcal{F}}(\varphi)^T \approx {\mathcal{F}}^\star(\varphi)^T = U_{\mathcal{F}} \left(P^T U_{\mathcal{F}}\right)^{-1} P^T \mathcal{F}(\varphi)^T,
\end{equation}
Transposing the previous expression gives
\begin{equation}
    {\mathcal{F}}(\varphi) \approx {\mathcal{F}}^\star(\varphi) = \mathcal{F}(\varphi)P
    \left(U_{\mathcal{F}}^TP\right)^{-1} U_{\mathcal{F}}^T.
\end{equation}
Substituting this expression into the scattered far-field equation (\ref{far-field equation}) yields
\begin{equation}
    e(\varphi) \approx \mathcal{F}^\star(\varphi) I^\star(\varphi) = \mathcal{F}(\varphi)P \left(U_{\mathcal{F}}^TP\right)^{-1} U_{\mathcal{F}}^T B \widetilde{I}(\varphi) \text{.}
\end{equation}
The term $\left(U_{\mathcal{F}}^TP\right)^{-1}U_{\mathcal{F}}^TB\in \mathbb{C}^{m \times r}$ does not depend on $\varphi$ and therefore can be precomputed. Also, $\mathcal{F}(\varphi)P \in \mathbb{C}^{1 \times m}$ can be obtained with ease by computing the components ${\wp_1}, \ldots, {\wp_m}$  of $\mathcal{F}(\varphi)$. Note that $m \ll n$.

The interpolation matrix \(P\) used in the far-field approximation is not recomputed. Instead, the same interpolation matrix previously obtained from the DEIM procedure applied to the excitation vector $V(\varphi)$ is used. Maxwell's equations satisfy reciprocity theorem. As a matter of fact, an antenna behaves the same both in transmission and reception. Thus, $P =\left[ e_{\wp_1}, e_{\wp_2}, \ldots, e_{\wp_m} \right] \in \mathbb{C}^{n \times m}$, where \(\wp_1,\ldots,\wp_m\) are the same interpolation indices for the excitation vector.

Instead of performing an additional SVD of far-field operator snapshots matrix, the basis $U_{\mathcal{F}} = \left[ u_{\mathcal{F},1}, u_{\mathcal{F},2}, \ldots, u_{\mathcal{F},m} \right] \in \mathbb{C}^{n \times m}$ is constructed from the modal coefficients already obtained in the SVD of the excitation snapshots matrix. Following the same expression as (\ref{lineal-combination u_i}), the \(i\)-th far-field mode is defined by replacing each excitation snapshot \(V(\varphi_j)\) with the corresponding transposed far-field operator snapshot \(\mathcal{F}(\varphi_j)^T\), while keeping the same expansion coefficients. As a result,
\begin{equation}
    u_{\mathcal{F},i} = \frac{1}{\sigma_i} \sum_{j=1}^{s} \mathcal{F}(\varphi_j)^T w_{j,i}, \qquad i=1,\ldots,m. 
\end{equation}

This procedure extends the DEIM strategy from the excitation term to the far-field evaluation. By reusing the interpolation indices and modal coefficients already obtained from the excitation snapshots, no additional SVD of far-field operator snapshots is required. Consequently, the scattered field can be computed directly from the
reduced current coefficients $\tilde{I}(\varphi) \in \mathbb{C}^{r \times 1}$, without reconstructing the MoM dimension electric current vector $I^\star(\varphi) \in \mathbb{C}^{n \times 1}$ for each incident angle ($r \ll n$). This paves the way to an efficient evaluation of the monostatic RCS regarless of the amount of incident angles required, since any unnecessary MoM dimension computation is conveniently avoided without sacrifying accuracy in the RCS results.
\section{Numerical Results}
\label{Sec-NumericalResults}

In this section, the proposed methodology is validated using several PEC targets with different geometrical and electrical characteristics. The whole procedure is fully automatic and non-intrusive: the user only needs to specify the simulation parameters, such as the angular range, frequency and number of angular samples, while the open-source solver Puma-EM \cite{PUMA-EM} is used as a black box to provide the MoM elements required to build the ROM. The geometries are meshed using \texttt{Gmsh} \cite{GeuR09}, and the electromagnetic problem is solved through the corresponding EFIE or CFIE formulation. All computations were carried out on a PC desktop with a 2.50-GHz $\text{11}^\text{th}$ Generation Intel Core i7-11700 processor and 16-GB memory.

For all test cases, the ROM was used to compute 4000 angular samples, whereas the MoM was evaluated at only 50 angles and used for reference comparison. The corresponding execution times are reported for the MoM and for two ROM evaluation strategies: the \emph{naive} approach, in which the MoM current approximation $I^\star(\varphi)$ is reconstructed before computing the far field, and the proposed approach, in which the DEIM approximation of the far-field operator is used to evaluate the RCS directly from the reduced current solution $\tilde{I}(\varphi)$.

\subsection{Cavity Problem}

\begin{figure}[tbp]
	\centering
	 \includegraphics[width=0.8\linewidth]{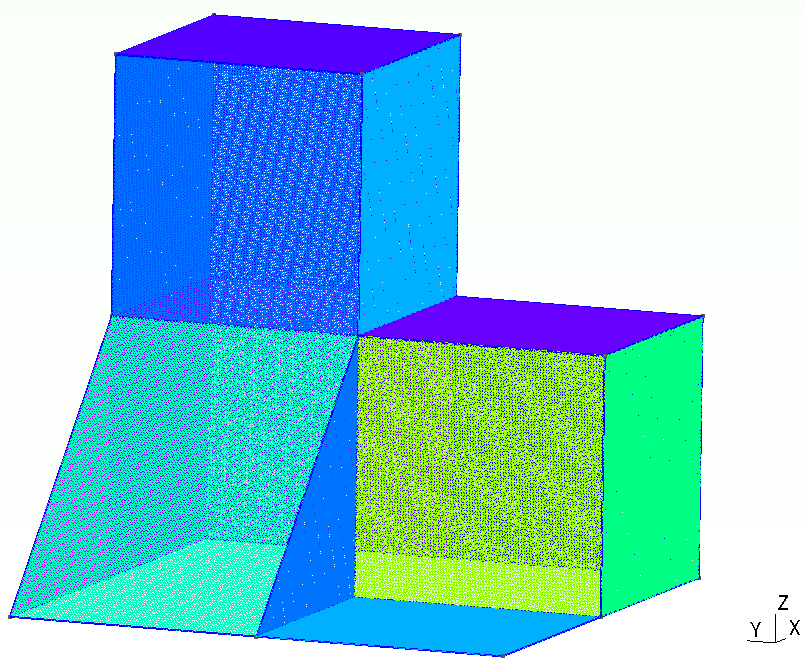}
	\caption{Cavity problem geometry and orientation.}
	\label{fig:cubiMesh}
\end{figure}

The first test case considers the PEC cavity shown in Fig.~\ref{fig:cubiMesh}. The structure is analyzed at 23 GHz, corresponding to an electrical size  of \(15.45\lambda\), for a monostatic angular sweep with  \(\theta=90^\circ\) and \(\varphi\in[90^\circ,180^\circ]\). The main simulation parameters are summarized in Table~\ref{tab:execution_params_Cubi}. The original MoM problem contains 338035  unknowns, whereas the greedy procedure achieves a ROM of dimension 39 and DEIM allows the excitation vector to be approximated by evaluating only 47 components. Regarding the computational cost, the MoM requires 335 min to compute 50 angular samples. In contrast, the ROM evaluates 4000 angles in 276 min when the full-order current approximation \(I^\star(\varphi)\) is reconstructed before the far-field computation. By using the proposed DEIM approximation of the far-field operator, the same 4000-point sweep is completed in 265 min.

\begin{figure}[tbp]
	\centering
	  \includegraphics[width=\linewidth]{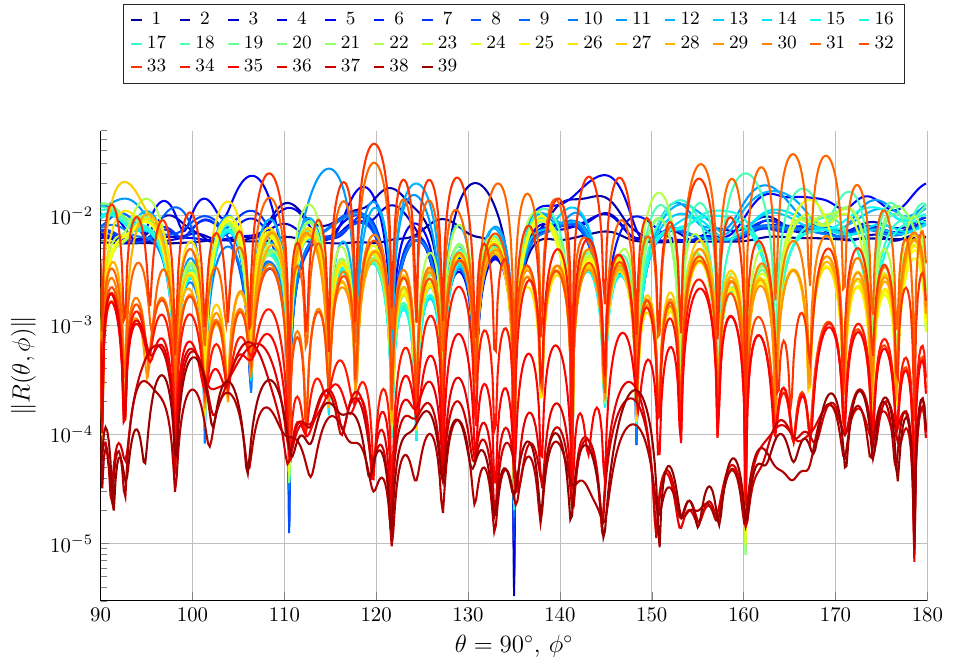}
	\caption{Residual error per iteration obtained in the greedy algorithm with the cavity problem.} 
	\label{fig:cubiResidualErrors}
\end{figure}

\begin{figure}[tbp]
	\centering
	 \includegraphics[width=\linewidth]{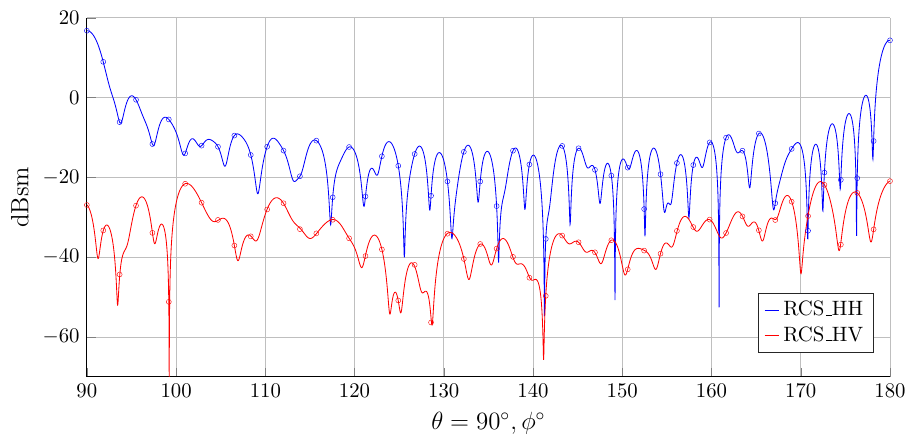}
	\caption{Monostatic RCS with horizontal polarization of the cavity problem. MoM~[$\circ$]. ROM~[--].} 
	\label{fig:cubiRCS}
\end{figure}

The evolution of the residual norm during greedy construction, following Algorithm~\ref{greedy}, is shown in Fig.~\ref{fig:cubiResidualErrors}. As new full-order solutions are incorporated into the reduced basis, the residual progressively decreases  over the entire angular interval, until the algorithm terminates after 39 basis vectors. Finally, Fig.~\ref{fig:cubiRCS} compares the monostatic RCS obtained with the MoM and the ROM. The MoM results, evaluated at 50 incidence angles, are represented by dots, whereas the continuous curves correspond to the 4000 angular samples computed with the reduced model. Both responses show excellent agreement over the complete angular range. This agreement confirms that, despite the drastic reduction in the dimension of the problem, the ROM is able to reproduce the behavior of the full-order model while preserving its accuracy. 
\begin{table}[tbp]
\caption{Test for the cavity problem}
\centering
\begin{tabular}{c c c}
\toprule
Angular Response & $\theta$ & 90º \\
           & $\phi$ & 90--180º \\
\hline
Frequency  & & 23 GHz (15.45$\lambda$) \\
\hline
           & ROM & 39 \\
Dimension  & MoM & 338035 \\
           & DEIM & 47 \\
\hline
        & $\tilde{I} \to t_{\mathrm{ROM},4000}$ & 265 min \\
Runtime & $I^\star \to t_{\mathrm{ROM},4000}$  & 276 min \\
        & $t_{\mathrm{MoM},50}$  & 335 min \\
\bottomrule
\end{tabular}
\label{tab:execution_params_Cubi}
\end{table}
\subsection{NASA Almond Problem}

\begin{figure}[tbp]
	\centering
	 \includegraphics[width=0.8\linewidth]{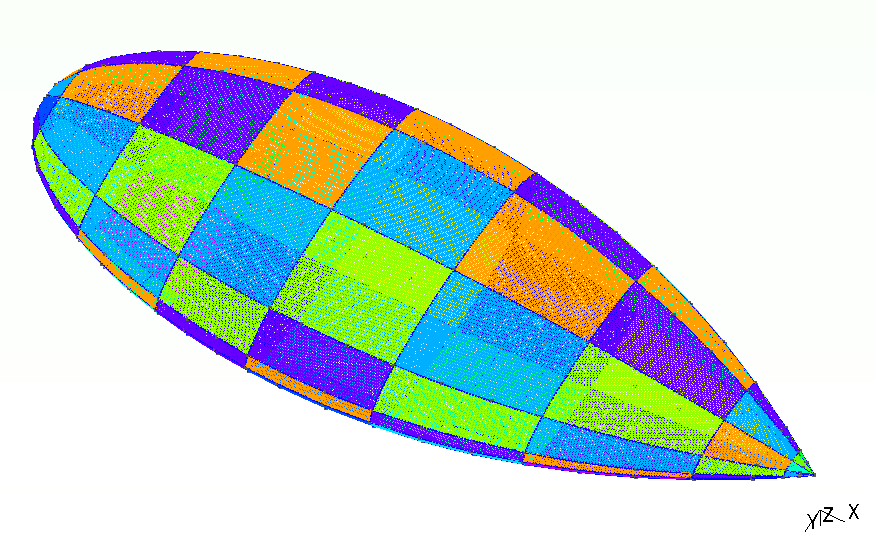}
	\caption{NASA almond geometry and orientation.}
	\label{fig:almondMesh}
\end{figure}

The second example considers the NASA almond, a widely used geometry in electromagnetic scattering analysis. The target is analyzed at 35 GHz, corresponding to an electrical size of \(29.57\lambda\), with 
\(\theta=90^\circ\) and an angular sweep in \(\varphi\) from 
\(-45^\circ\) to \(45^\circ\). The geometry and the main simulation
parameters are shown in Fig.~\ref{fig:almondMesh} and Table~\ref{tab:execution_params_Almond_Bottom}, respectively. This example involves a full-order problem with 222624 MoM unknowns. Nevertheless, the greedy procedure reduces the system to a ROM of dimension 33, while the DEIM approximation requires only 45 interpolation components. Regarding the computational cost, the MoM requires 20 min to evaluate 50 angular samples, whereas the ROM computes 4000 samples in 15 min when the DEIM approximation of the far-field operator is employed.

\begin{figure}[tbp]
	\centering
	 \includegraphics[width=\linewidth]{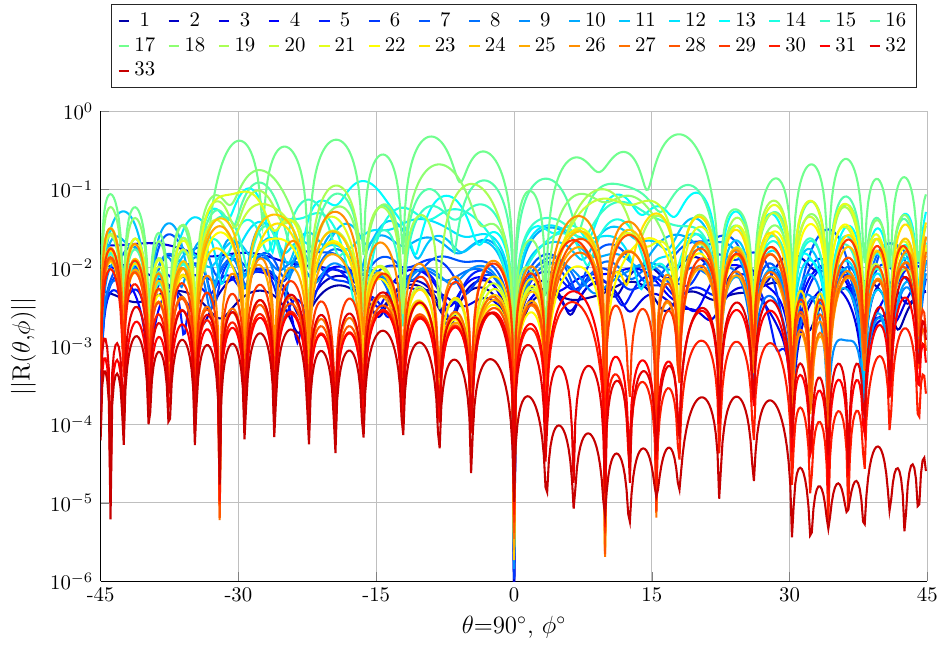}
	\caption{Residual error per iteration obtained in the greedy algorithm with the NASA almond.} 
	\label{fig:almondBottomResidualErrors}
\end{figure}

\begin{figure}[tbp]
	\centering
	 \includegraphics[width=\linewidth]{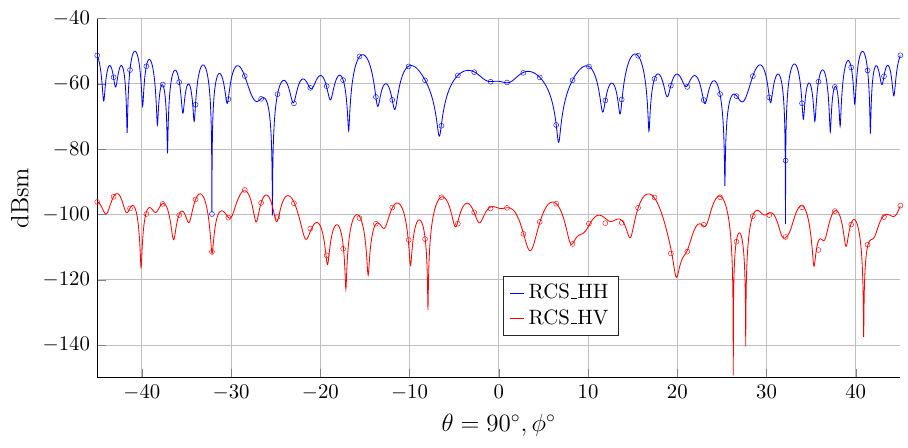}
	\caption{Monostatic RCS with horizontal polarization of the NASA almond. MoM~[$\circ$]. ROM~[--].} 
	\label{fig:almondBottomRCS}
\end{figure}

The evolution of the residual during the greedy process is shown in Fig.~\ref{fig:almondBottomResidualErrors}, illustrating how the reduced basis is progressively enriched.The monostatic RCS obtained with the reduced and full-order models is compared in Fig.~\ref{fig:almondBottomRCS}.
\begin{table}[tbp]
\caption{Test for the NASA Almond}
\centering
\begin{tabular}{c c c}
\toprule
Angular Response & $\theta$ & 90º \\
           & $\phi$ & -45--45º \\
\hline
Frequency  & & 35 GHz (29.57$\lambda$) \\
\hline
           & ROM & 33 \\ 
Dimension  & MoM & 222624 \\
           & DEIM & 45 \\
\hline
        & $\tilde{I} \to t_{\mathrm{ROM},4000}$ & 15 min \\
Runtime & $I^\star \to t_{\mathrm{ROM},4000}$  & 27 min \\
        & $t_{\mathrm{MoM},50}$  & 20 min \\

\bottomrule
\end{tabular}
\label{tab:execution_params_Almond_Bottom}
\end{table}
\subsection{Turbine Inlet Model Problem}

\begin{figure}[tbp]
	\centering
	 \includegraphics[width=0.7\linewidth]{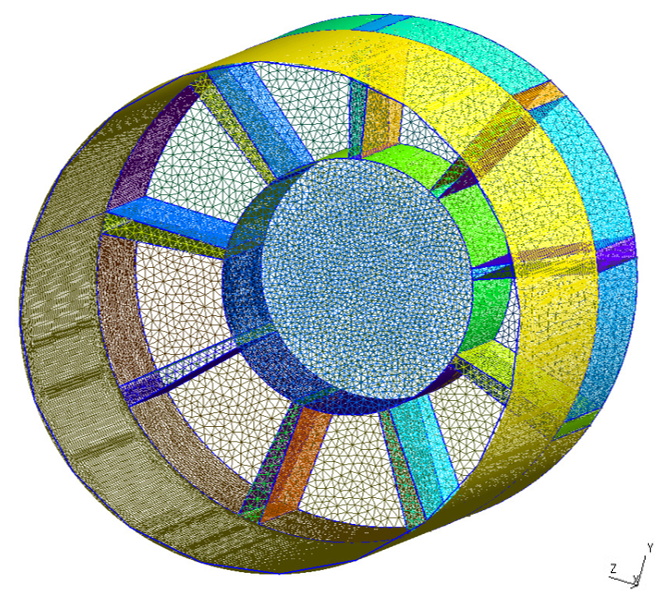}
	\caption{Turbine inlet model geometry and orientation.}
	\label{fig:turbineMesh}
\end{figure}

The third example considers a turbine inlet model geometry, shown in Fig.~\ref{fig:turbineMesh}. The turbine inlet model dimensions can be found in \cite{Jin2003}. The analysis is performed at 5.5 GHz, which corresponds to an electrical size of \(5.61\lambda\), with \(\theta=90^\circ\) and \(\varphi\in[0^\circ,180^\circ]\). The corresponding simulation parameters are summarized in Table~\ref{tab:execution_params_Turbine}. In this case, the original MoM problem contains 89942 unknowns, whereas the greedy algorithm constructs a reduced model of dimension 35 and the DEIM approximation requires 41 interpolation components. From a computational point of view, the MoM requires 106 min to evaluate only 50 angular samples, whereas the ROM, using the DEIM approximation of the far-field operator, completes the evaluation of 4000 angles in 74 min. 

\begin{figure}[tbp]
	\centering
	\includegraphics[width=\linewidth]{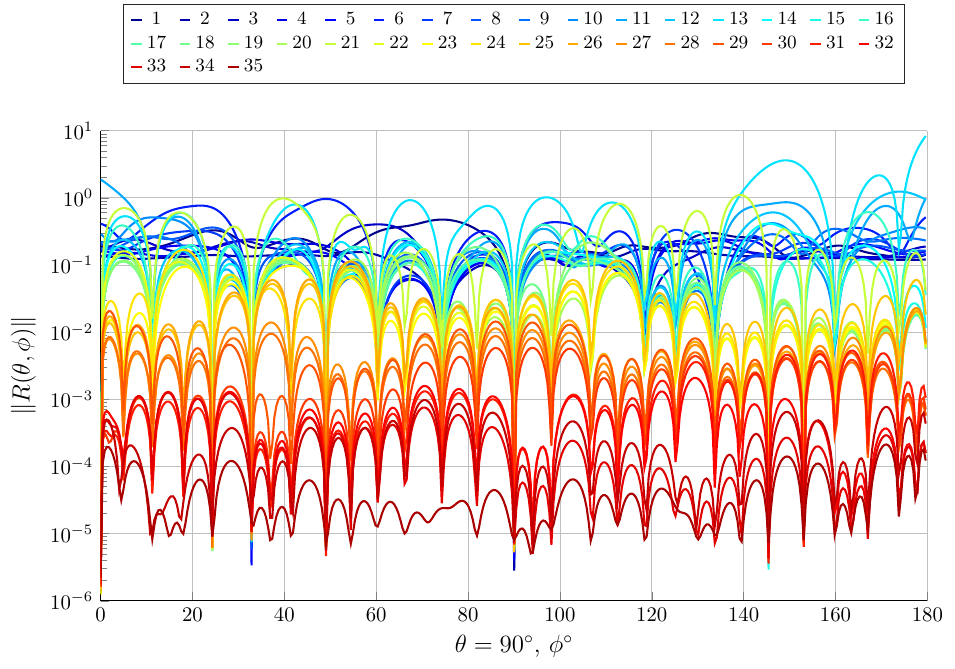}
	\caption{Residual error per iteration obtained in the greedy algorithm with the turbine inlet model.} 
	\label{fig:turbineResidualErrors}
\end{figure}

\begin{figure}[tbp]
	\centering
	\includegraphics[width=\linewidth]{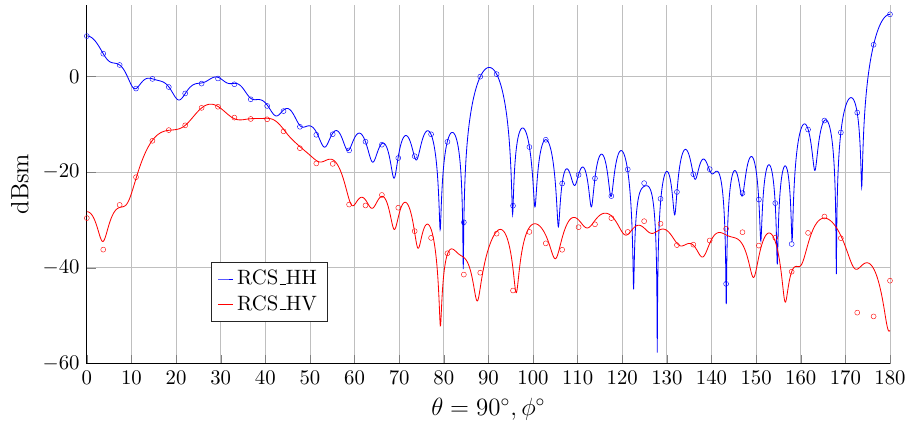}
	\caption{Monostatic RCS with horizontal polarization of the turbine inlet model. MoM~[$\circ$]. ROM~[--].} 
	\label{fig:turbineRCS}
\end{figure}

The evolution of the residual during the construction of the reduced basis is shown in Fig.~\ref{fig:turbineResidualErrors} and the RCS comparison is presented in Fig.~\ref{fig:turbineRCS}. Overall, the ROM reproduces the behavior of the full-order response across the angular range, although it can be observed that the results do not match at certain points. A possible explanation is the presence of numerical errors in the full-order current solutions generated by the MoM solver.
Since these solutions are used as snapshots during the construction of the reduced basis, any inaccuracy present in the MoM data can also be inherited by the ROM. Despite these localized differences, the reduced model still reproduces the main characteristics and the overall angular
evolution of the RCS.
\begin{table}[tbp]
\caption{Test for the Turbine Inlet Model}
\centering
\begin{tabular}{c c c}
\toprule
Angular Response & $\theta$ & 90º \\
           & $\phi$ & 0--180º \\
\hline
Frequency  & & 5.5 GHz (5.61$\lambda$) \\
\hline
           & ROM & 35 \\
Dimension  & MoM & 89942 \\
           & DEIM & 41 \\
\hline
        & $\tilde{I} \to t_{\mathrm{ROM},4000}$ & 74 min \\
Runtime & $I^\star \to t_{\mathrm{ROM},4000}$  & 77 min \\
        & $t_{\mathrm{MoM},50}$  & 106 min \\

\bottomrule
\end{tabular}
\label{tab:execution_params_Turbine}
\end{table}
\subsection{F-16 Aircraft Model Problem}

\begin{figure}[tbp]
	\centering
	 \includegraphics[width=1.0\linewidth]{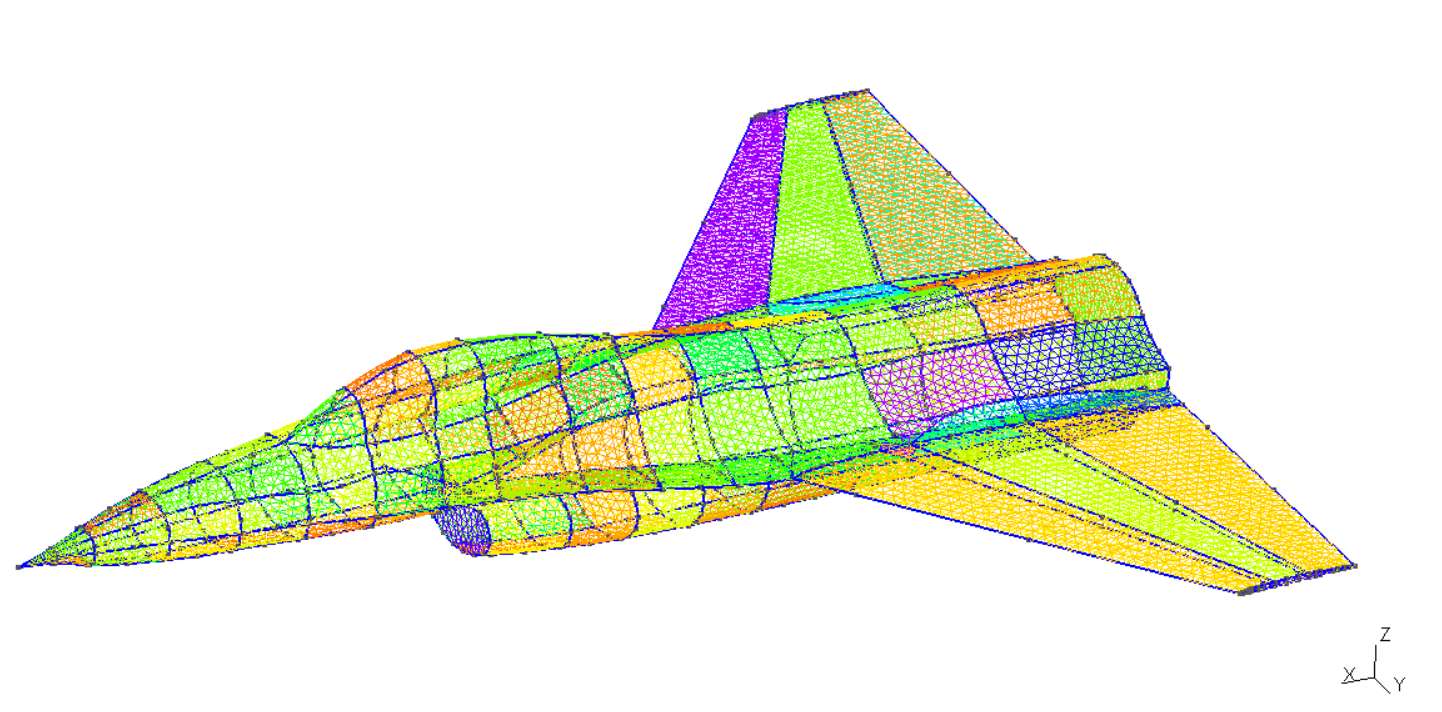}
	\caption{F-16 aircraft model geometry and orientation.}
	\label{fig:f16Mesh}
\end{figure}

The last example considers an F-16 aircraft model, shown in Fig.~\ref{fig:f16Mesh}. The target is analyzed at 500 MHz, which
corresponds to an electrical size of \(19.65\lambda\), for \(\theta=90^\circ\) and \(\varphi\in[0^\circ,90^\circ]\). The main simulation parameters are summarized in Table~\ref{tab:execution_params_f16}. This case leads to a full-order MoM problem with 125721 unknowns, which is reduced to a system of dimension 41, while 45 components are retained in the DEIM
approximation. In terms of computational effort, the MoM requires 30 min to evaluate 50 incidence angles, whereas the proposed ROM with DEIM applied to the far-field operator completes a 4000-point angular sweep in 27 min. This allows a substantially finer angular resolution to be obtained without the cost associated with performing thousands of full-order simulations.

\begin{figure}[tbp]
	\centering
	\includegraphics[width=\linewidth]{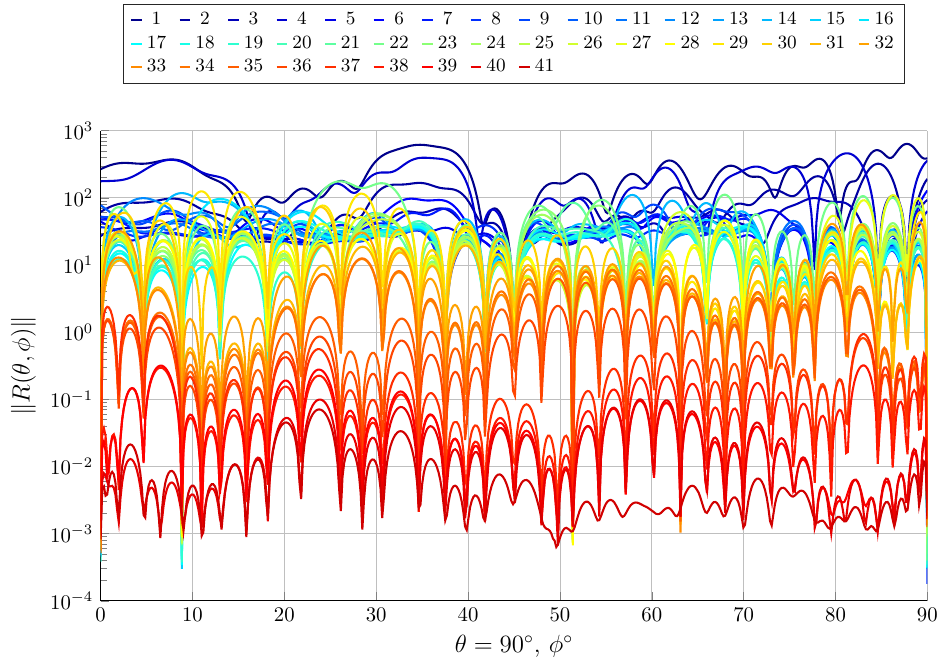}
	\caption{Residual error per iteration obtained in the greedy algorithm with the F-16 aircraft model.} 
	\label{fig:f16ResidualErrors}
\end{figure}

\begin{figure}[tbp]
	\centering
	\includegraphics[width=\linewidth]{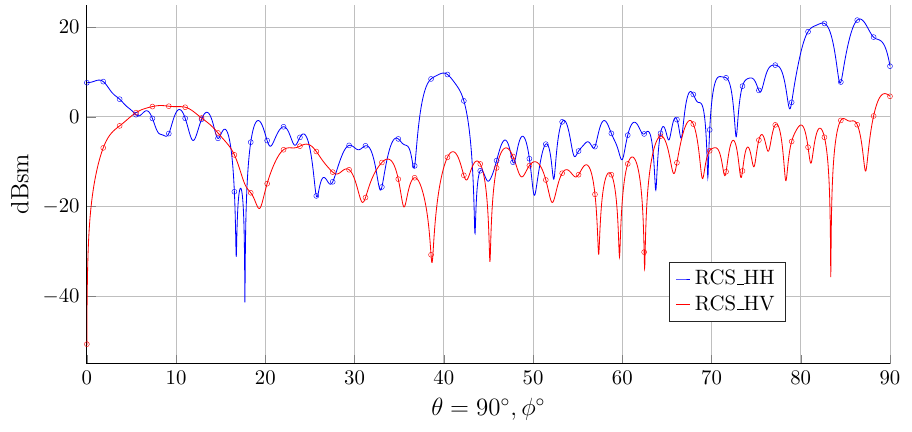}
	\caption{Monostatic RCS with horizontal polarization of the F-16 aircraft model. MoM~[$\circ$]. ROM~[--].} 
	\label{fig:f16RCS}
\end{figure}

Fig.~\ref{fig:f16ResidualErrors} shows the evolution of the residual throughout the greedy iterations. Despite the geometrical complexity of the target and the large reduction in the system dimension, the monostatic RCS comparison in Fig.~\ref{fig:f16RCS} shows a good agreement between the MoM reference samples and the continuous response obtained with the ROM, which is able to reproduce the main features of the angular response.
\begin{table}[tbp]
\caption{Test for the F-16 Aircraft Model}
\centering
\begin{tabular}{c c c}
\toprule
Angular Range & $\theta$ & 90º \\
           & $\phi$ & 0--90º \\
\hline
Frequency  & & 500 MHz (19.65$\lambda$) \\
\hline
           & ROM & 41 \\
Dimension  & MoM & 125721 \\
           & DEIM & 45 \\
\hline
    & $\tilde{I} \to t_{\mathrm{ROM},4000}$ 
    & 27  min \\
Runtime  & $I^\star \to t_{\mathrm{ROM},4000}$  & 37 min \\
           & $t_{\mathrm{MoM},50}$ & 30 min \\
           
\bottomrule
\end{tabular}
\label{tab:execution_params_f16}
\end{table}
\section{Conclusions}
\label{Sec-Conclusions}
A reduced-order modeling strategy for fast full-wave
monostatic RCS evaluation over dense angular sweeps has been presented. The proposed methodology combines RBM and DEIM within a non-intrusive framework, using an open-source integral equation solver as a black box. In this way, the full-order MoM solver is only required at a limited number of selected incidence angles, while the final angular response is obtained from a low-dimensional reduced model.

The RBM approximation reduces the dimension of the current solution space by constructing a compact basis from representative full-order solutions selected through a greedy algorithm. The residual error estimator provides an efficient criterion to guide this construction and identify the angles where additional information is required. In parallel, DEIM reduces the cost associated with the evaluation of the parameter-dependent excitation vector by requiring only a small number of selected components.

Furthermore, the DEIM strategy has been extended to the
far-field evaluation. This allows the scattered field and the monostatic RCS to be computed directly from the reduced current coefficients, avoiding the reconstruction of the full-order current vector. As a result, the online stage remains within the reduced space, improving the efficiency of the overall procedure.
\section*{Acknowledgement}
This work has been developed within the framework of the activities of the project \emph{Plasma reconfigUrabLe metaSurface tEchnologies} (PULSE), funded by the European Innovation Council under the EIC Pathfinder Open 2022 program (protocol number 101099313). The project website is: \url{https://www.pulse-pathfinder.eu/}.
\bibliographystyle{sty/IEEEtran}
\bibliography{bibliography/references}

@article{GeuR09,
  title={Gmsh: A {3-D} finite element mesh generator with built-in pre-and post-processing facilities},
  author={Geuzaine, Christophe and Remacle, Jean-Fran{\c{c}}ois},
  journal={Int. J. Numer. Methods Eng.},
  volume={79},
  number={11},
  pages={1309--1331},
  year={2009},
  publisher={Wiley Online Library}
}

@article{Silveira2014Reduced-OrderModels,
  author        = "A. Hochman and J. {Fern\'{a}ndez Villena} and A. G. Polimeridis and L. M. Silveira and J. K. White and L. Daniel",
  title         = "Reduced-Order Models for Electromagnetic Scattering Problems",
  journal={{IEEE} Trans. Antennas Propagat.},
  volume        = "62",
  number	= "6",
  pages         = "3150--3162",
  year          = "2014"
}

@article{Edlinger2014APosteriori,
  author        = "Yves Konkel and Ortwin Farle and Alexander Sommer and Stefan Burgard and Romanus Dyczij-Edlinger",
  title         = "A Posteriori Error Bounds for Krylov-Based Fast Frequency
Sweeps of Finite-Element Systems",
  journal       = {{IEEE} Trans. Magn.},
  volume        = "50",
  number	= "2",
  pages         = "441-444",
  year          = "2014"
}

@book{QuarteroniManzoniRBM,
  author        = {A. Quarteroni and A. Manzoni and F. Negri},
  title         = {Reduced Basis Methods for Partial Differential Equations},
  publisher     = {Springer},
  address       = {Cham, Switzerland},
  year          = {2016}
}

@book{HesthavenRozzaRBM,
  author        = {Hesthaven, Jan S and Rozza, Gianluigi and Stamm, Benjamin},
  title         = {Certified reduced basis methods for parametrized partial differential equations},
  publisher     = {Springer},
  address       = {Berlin, Germany},
  year          = {2016}
}

@article{benner2015survey,
  title={A survey of projection-based model reduction methods for parametric dynamical systems},
  author={Benner, Peter and Gugercin, Serkan and Willcox, Karen},
  journal={SIAM review},
  volume={57},
  number={4},
  pages={483--531},
  year={2015},
  publisher={SIAM},
}

@article{de2009reliable,
  title={Reliable fast frequency sweep for microwave devices via the reduced-basis method},
  author={de la Rubia, Valent\'in and Razafison, Ulrich and Maday, Yvon},
  journal={{IEEE} Trans. Microwave Theory Tech.},
  volume={57},
  number={12},
  pages={2923--2937},
  year={2009},
  publisher={IEEE}
}

@article{MattRCS2010,
  author={Peng, Zhen and Stephanson, Mattew B. and Lee, Jin-Fa},
  journal={{IEEE} Trans. Antennas Propagat.},
  title={Fast Computation of Angular Responses of Large-Scale Three-Dimensional Electromagnetic Wave Scattering}, 
  year={2010},
  volume={58},
  number={9},
  pages={3004-3012},
  doi={10.1109/TAP.2010.2052579}}

@misc{PUMA-EM,
    title = {{Puma-EM}},
    howpublished = {Available: \url {https://github.com/Gjacquenot/Puma-EM}} 
}

@article{chaturantabut2010nonlinearDEIM,
  title={Nonlinear model reduction via discrete empirical interpolation},
  author={Chaturantabut, Saifon and Sorensen, Danny C},
  journal={SIAM Journal on Scientific Computing},
  volume={32},
  number={5},
  pages={2737--2764},
  year={2010},
  publisher={SIAM}
}

@article{monje2020electricRBMEFIE,
  title={Electric field integral equation fast frequency sweep for scattering of nonpenetrable objects via the reduced-basis method},
  author={Monje-Real, Alberto and de la Rubia, Valent{\'\i}n},
  journal={{IEEE} Trans. Antennas Propagat.},
  volume={68},
  number={8},
  pages={6232--6244},
  year={2020},
  publisher={IEEE}
}

@article{fares2011reducedRBMEFIE,
  title={The reduced basis method for the electric field integral equation},
  author={Fares, M and Hesthaven, Jan S and Maday, Yvon and Stamm, Benjamin},
  journal={Journal of Computational Physics},
  volume={230},
  number={14},
  pages={5532--5555},
  year={2011},
  publisher={Elsevier}
}

@article{chellappa2023infsup,
  author={Chellappa, Sridhar and Feng, Lihong and de la Rubia, Valentín and Benner, Peter},
  journal={{IEEE} Trans. Microwave Theory Tech.}, 
  title={Inf-Sup-Constant-Free State Error Estimator for Model Order Reduction of Parametric Systems in Electromagnetics}, 
  year={2023},
  volume={71},
  number={11},
  pages={4762-4777},
  doi={10.1109/TMTT.2023.3288642}
  }

@article{delaRubia2022fastaposteriori,
  author={de la Rubia, Valentin and Chellappa, Sridhar and Feng, Lihong and Benner, Peter},
  journal={{IEEE} Trans. Microwave Theory Tech.},
  title={Fast A Posteriori State Error Estimation for Reliable Frequency Sweeping in Microwave Circuits via the Reduced-Basis Method}, 
  year={2022},
  volume={70},
  number={11},
  pages={5172-5184},
  doi={10.1109/TMTT.2022.3203726}
  }

@article{morRewLM15,
  title={A goal-oriented error estimator for reduced basis method modeling of microwave devices},
  author={Rewie{\'n}ski, Micha{\l} and Lamecki, Adam and Mrozowski, Micha{\l}},
  journal={IEEE Microw. Wirel. Compon. Lett.},
  volume={25},
  number={4},
  pages={208--210},
  year={2015},
  publisher={IEEE}
}

@ARTICLE{JinFaRCS2023,
  author={Lee, Chung Hyun and Kotulski, Joseph D. and Dang, Vinh Q. and Lee, Jin-Fa},
  journal={{IEEE} Trans. Antennas Propagat.}, 
  title={An Efficient and Error-Controllable Angular Sweep Algorithm for Electromagnetic Wave Scattering and its Analysis}, 
  year={2023},
  volume={71},
  number={3},
  pages={2613-2625},
  doi={10.1109/TAP.2022.3233426}}

@INPROCEEDINGS{LandesaRCS2025,
  author={Gomez-Rodriguez, A. and Landesa, L. and Martin, V. F. and Taboada, J. M. and Obelleiro, F.},
  booktitle={2025 19th European Conference on Antennas and Propagation (EuCAP)}, 
  title={Self-Correcting Interpolation Technique for Multi-Source Electromagnetic Scattering Problems}, 
  year={2025},
  volume={},
  number={},
  pages={1-4},
  doi={10.23919/EuCAP63536.2025.10999545}}

@article{lee2023efficient,
  title={An efficient and error-controllable angular sweep algorithm for electromagnetic wave scattering and its analysis},
  author={Lee, Chung Hyun and Kotulski, Joseph D and Dang, Vinh Q and Lee, Jin-Fa},
  journal={{IEEE} Trans. Antennas Propagat.},
  volume={71},
  number={3},
  pages={2613--2625},
  year={2023},
  publisher={IEEE}
}

@article{Jin2003,
  author={Jian-Ming Jin and Jian Liu and Zheng Lou and Liang, C.S.T.},
  journal={{IEEE} Trans. Antennas Propagat.},
  title={A fully high-order finite-element simulation of scattering by deep cavities}, 
  year={2003},
  volume={51},
  number={9},
  pages={2420-2429},
  doi={10.1109/TAP.2003.816354}}

@article{Rius2026,
  author={Heldring, Alex and Ubeda, Eduard and Rius, Juan M.},
  journal={{IEEE} Trans. Antennas Propagat.},
  title={Fast Direct High-Frequency MoM Solver Using Butterfly Algebra}, 
  year={2026},
  volume={74},
  number={1},
  pages={703-712},
  doi={10.1109/TAP.2025.3627867}}

@article{Michielssen2017,
  author={Guo, Han and Liu, Yang and Hu, Jun and Michielssen, Eric},
  journal={{IEEE} Trans. Antennas Propagat.},
  title={A Butterfly-Based Direct Integral-Equation Solver Using Hierarchical LU Factorization for Analyzing Scattering From Electrically Large Conducting Objects}, 
  year={2017},
  volume={65},
  number={9},
  pages={4742-4750},
  doi={10.1109/TAP.2017.2727511}}

@article{Schuster2012,
  author={Schroder, Arne and Brüns, Heinz-D. and Schuster, Christian},
  journal={{IEEE} Trans. Antennas Propagat.},
  title={A Hybrid Approach for Rapid Computation of Two-Dimensional Monostatic Radar Cross Section Problems With the Multilevel Fast Multipole Algorithm}, 
  year={2012},
  volume={60},
  number={12},
  pages={6058-6061},
  doi={10.1109/TAP.2012.2209858}}

@article{Sheng2014,
  author={Pan, Xiao-Min and Sheng, Xin-Qing},
  journal={{IEEE} Trans. Antennas Propagat.},
  title={Accurate and Efficient Evaluation of Spatial Electromagnetic Responses of Large Scale Targets}, 
  year={2014},
  volume={62},
  number={9},
  pages={4746-4753},
  doi={10.1109/TAP.2014.2334356}}
\end{document}